\documentclass[a4paper,11pt]{article}
\pdfoutput=1 

\usepackage{jheppub}

\title{\boldmath
 Conformal symmetry limit of
QED and QCD and     identities  between
perturbative contributions to
deep-inelastic scattering  sum rules}

\author{A.L. Kataev}

\affiliation{Institute for Nuclear Resrach of the Russian
Academy of Sciences,\\
117312 Moscow, Russia}

\emailAdd{kataev@ms2.inr.ac.ru}

\abstract{Conformal symmetry-based    relations
between concrete perturbative QED and QCD  approximations  for
the     Bjorken ,  the     Ellis-Jaffe sum rules
of  polarized lepton- nucleon  deep-inelastic scattering (DIS),
the   Gross-Llewellyn Smith sum rules of neutrino-nucleon DIS,
and for  the Adler functions of   axial-vector and vector channels
are derived.
They result from  the   application of the operator product
expansion  to    three  triangle   Green functions,
constructed from  the
non-singlet axial-vector,  and two  vector currents, the
 singlet axial-vector and two  non-singlet vector currents and
the non-singlet  axial-vector,  vector and
singlet vector currents     in the   limit,   when the  conformal
symmetry  of the  gauge  models with fermions is considered  unbroken.
We specify the perturbative conditions for
this  symmetry to be  valid in the case of   the   $U(1)$ and $SU(N_c)$  models.
The
all-order perturbative    identity following from the   conformal invariant limit  between
the concrete  contributions to
the Bjorken,  the   Ellis-Jaffe  and the   Gross-Llewellyn
Smith sum rules  is    proved.
The analytical and numerical $O(\alpha^4)$ and $O(\alpha_s^2)$
conformal symmetry based
approximations    for    these
sum rules and for   the    Adler function of the  non-singlet
vector currents  are summarized.  Possible
theoretical  applications of the results  presented   are discussed.\\

}

\begin{document}
\maketitle
\flushbottom

\section{Introduction .}
\label{sec:intro}

The concepts of  scale-invariance and of    conformal symmetry
(CS) play an
important role in  studies of perturbative
approximations for various three-point and two-point Green functions
of  the  renormalized massless quantum  gauge models
with fermions. It   was proved  in the coordinate space-time representation,
that in  the CS  limit of  these models,
which is realized  when  coupling constants are considered  as  non-renormalized   bare  parameters
(i.e. when $\alpha=\alpha^B$ or $\alpha_s=\alpha_s^{B}$),
the multiloop expression for the AVV
three-point  Green function  of the   flavour
non-singlet(NS)
axial-vector and two  vector currents  coincides with the lowest-order
 1-loop  triangle graph  ~\cite{Schreier:1971um}.
In the momentum space-time representation the result
of ref.~\cite{Schreier:1971um} was rewritten in ref.~\cite{Gabadadze:1995ei}
as
\begin{equation}
\label{Tri}
T_{\mu\alpha\beta}^{cab}(p,q)=i\int<0|T A_{\mu}^{c}(y)
V_{\alpha}^a(x)V_{\beta}^b(0)|0>e^{ipx+iqy}dxdy=
d_R^{cab}\Delta_{\mu\alpha\beta}^{1-l}(p,q)~~~.
\end{equation}
Here
$A_{\mu}^{c}(y)=\overline{\psi}(y)\gamma_{\mu}(\lambda^{c}/2)\gamma_{5}\psi(y)$
and
$V_{\alpha}^{a}(x)=
\overline{\psi}(x)\gamma_{\alpha}(\lambda^{a}/2)\psi(x)$ are
the NS axial-vector and vector  currents, $d_R^{cab}$ is the
symmetric  structure constant of the $SU(N_c)$ group with its 
generators $(\lambda^{a}/2)$, $(\lambda^{b}/2)$,  $(\lambda^{c}/2)$ 
defined in the representation $R$ of the
related Lie algebra   and
$\Delta_{\mu\alpha\beta}^{1-l}(p,q)$ is the 1-loop  contribution
to  the triangle  Green function.
In ref. \cite{Erlich:1996mq} the validity of
this non-renormalization  property  was explicitly demonstrated
at the 2-loop level
using  differential regularization and    differential
renormalization prescriptions, proposed in ref.~\cite{Freedman:1991tk}.
For the   dimensional regularization ~\cite{'tHooft:1972fi},
~\cite{Bollini:1972ui},~\cite{Bollini:1972bi},~\cite{Ashmore:1972uj},
~\cite{Cicuta:1972jf}   and
in the $\overline{MS}$-scheme ~\cite{Bardeen:1978yd},
formulated in more detail in ~\cite{Kataev:1988sq},
the cancellation of all 2-loop internal  contributions
to the AVV three-point function  were
rediscovered in ref.  ~\cite{Jegerlehner:2005fs}.
The agreement of this  result with the outcomes of 2-loop
calculations,  
performed in  ref.~\cite{Erlich:1996mq} using
differential regularization and renormalization approaches,
is not accidental.  Indeed, 
the differential renormalization can be straightforwardly
related to the dimensional regularization and the $\overline{MS}$-scheme
\cite{Smirnov:1994km}.

In the present  work  we   consider   two extra
AVV three-point Green functions, which are closely related to the one
of eq.~\eqref{Tri}. The first of them is
constructed from the  singlet (SI)  axial-vector  and
two NS vector currents, namely
\begin{equation}
\label{tw}
\tilde{T}_{\mu\alpha\beta}^{ab}(p,q)=i\int<0|TA_{\mu}(y)
V_{\alpha}^a(x)V_{\beta}^b(0)|0>e^{ipx+iqy}dxdy~~~.
\end{equation}
where
$A_{\mu}(y)=\overline{\psi}(y)\gamma_{\mu}\gamma_5\psi(y)$
is the SI axial-vector fermion current.
The properties of this  Green function were  investigated   previously
in ref.~\cite{Crewther:1973zb} within the
deeply investigated
finite QED program (see e.g. 
refs.~\cite{Johnson:1967pk,Adler:1972in}).
This program had the  aim  to find out  whether a  non-trivial
ultraviolet zero may exist in the RG $\beta$-function of the perturbative
quenched QED (pqQED) model ( which will  be defined in Section 2) or in the
QED Gell-Mann-Low  function $\Psi(\alpha)$ ~\cite{Johnson:1973pd}, which as
clarified in the review of ref. ~\cite{Vladimirov:1979ib}
is identical to the QED $\beta$-function in the momentum subtractions
scheme (see  e.g.   ref.~\cite{Gorishnii:1990kd} for the detailed explanation ).
Using the  methods of   the finite QED program, it   was shown  in 
ref.\cite{Adler:1972in}, that  if an  ultraviolet zero of
$\Psi(\alpha)$ exists, it should be a   zero of infinite order. Now we know
that this feature is not realized in perturbative QED.
However, the important understanding of the basic features of the methods and
the symmetries, gained in the process of investigations of the
of the finite QED program, remain important today.
Among them is the notion of the CS limit of QED. The 
consequences of the possible  applications of the  limit to  the
perturbative expression of eq.~\eqref{tw} were not considered in
ref.\cite{Crewther:1973zb}.
 The concrete
analytical high-order perturbative results,
which follow from this limit, were obtained only recently~\cite{Kataev:2010tm}.
The second  three-point Green function we  will be interested in
is  constructed from the  NS axial-vector , the  NS
vector and SI vector  fermion currents with $N_F$ number of fermions,
namely
\begin{equation}
\label{3}
\tilde{\tilde{T}}_{\mu\alpha\beta}^{ab}(p,q)=i\int<0|TA_{\mu}^{a}(y)
V_{\alpha}^{b}(0)V_{\beta}(y)|0>e^{iqx-ipy}dxdy~~~.
\end{equation}
Here $V_{\beta}(0)=N_F\overline{\psi}_i(0)\gamma_{\alpha}\psi_i(0)$  is the SI
vector quark current with $N_F$ fermions of unit charge. Theoretical consequences,
which will be obtained from  these
two AVV  functions of eq.~\eqref{tw} and eq.~\eqref{3} are new  and were not
published in the
regular journal, though some of them were already 
discussed by the author (see ~\cite{Kataev:1996ce},
~\cite{Kataev:2012kc}).

The most important  result, derived from  eq.~\eqref{Tri}
in the conformally  invariant limit,
is the relation
between $\pi\rightarrow\gamma\gamma$ decay constant and
the product of the Bjorken sum rule of the
polarized deep-inelastic scattering (DIS) and the
Adler functions, related to the total cross-section
of the $e^+e^-$-annihilation to hadrons process,
both evaluated in the Born approximation ~\cite{Crewther:1972kn}.
Other relations, which follow from the application of the
operator product expansion (OPE) approach
to the same  AVV Green function in different kinematic regimes,
were derived in  ref.~\cite{Adler:1973kz}.
In this work the basic Crewther relation of ref.~\cite{Crewther:1972kn}
was generalized to the
the level of the  $O(\alpha^2)$ corrections  within
finite QED program.

In the case of the  $SU(N_c)$ gauge model with fermions
the most important understanding of the properties
of the  perturbative series
in the case when the
CS exist and when it is  violated  by the $\overline{MS}$ renormalization procedure
 was achieved in  ref.~\cite{Broadhurst:1993ru}. In this work
the generalized QCD  Crewther  relation   between the analytical
$\overline{MS}$-scheme
$O(\alpha_s^3)$-approximations
for  the Bjorken  polarized (Bjp)  sum rule ,
the  Gross-Llewellyn Smith (GLS) sum rule ~\cite{Larin:1991tj} and the similar approximation for
the $e^+e^-$-annihilation Adler Function, evaluated in
refs.~\cite{Gorishnii:1990vf,Surguladze:1990tg}
\footnote{In  numerical form
the result of ref.\cite{Gorishnii:1990vf} was published in
ref.~\cite{Gorishnii:1991se}.}
and independently confirmed  in ref.  ~\cite{Chetyrkin:1996ez},
was  discovered. The factorization  of the
the 2-loop RG $\beta$-function of the $SU(N_c)$-model was revealed in the generalized
$\overline{MS}$-scheme Crewther relation at the $\alpha_s^3$-level.

The   analytical calculations
of  the $\alpha_s^4$-corrections to
the Bjp sum rule~\cite{Baikov:2010je},  the GLS  sum
rule, the $D$-function in the NS and vector channels allowed the authors
of refs. ~\cite{Baikov:2010je,Baikov:2012zn} 
to demonstrate explicitly  the existence of the closed
$\overline{MS}$-scheme approximations   for    the generalized   QCD
Crewther relations  ~\cite{Broadhurst:1993ru}    with   the factorized 3-loop 
RG  $\beta$-function,
analytically evaluated in ref.~\cite{Tarasov:1980au} and confirmed in
ref.~\cite{Larin:1993tp}. The validity of  the
$\overline{MS}$-scheme   relation
for the Bjp sum rule to all orders of perturbation theory
was  considered in  momentum space ~\cite{Gabadadze:1995ei} and then
proved in the coordinate
space-time representation ~\cite{Crewther:1997ux,Braun:2003rp} without specifying the form
of high order perturbative QCD corrections.
The  additional   SI-type $\alpha_s^4$ contributions to
the $O(\alpha_s^4)$ NS expression for the Bjp
sum rule \cite{Baikov:2010je}
 were obtained recently ~\cite{Larin:2013yba}
\footnote{The results are   not yet verified   by direct analytical calculations.}
from the equality of the generalizations of the Crewther relations for the
product of the coefficient function of the Bjp sum rule and the NS $D$-function and the
product of the coefficient functions of the GLS sum rule and the $D$-function of vector
currents. This equality was proposed in ref.~\cite{Kataev:1996ce} and published 
in ref.~\cite{Kataev:2011im} without
taking into account SI-type $O(\alpha_s^4)$-corrections to the Bjp sum rule. 
Note, that the Crewther relation for the GLS sum rule was analysed  in the Born QED approximation 
~\cite{Adler:1973kz}, in the $O(a_s^3)$ and $O(a_s^4)$ $SU(N_c)$ studies of ref.~~\cite{Broadhurst:1993ru} and ref.~\cite{Baikov:2012zn}, and in the 
general theoretical perturbative QCD analysis of 
refs.~\cite{Crewther:1997ux,Braun:2003rp} as well.

In this work   we derive new   perturbative all-order
identities  between the  concrete approximations for  the
coefficient functions of the  Bjp,  Ellis-Jaffe (EJ) and
the GLS sum rules. In the CS  limit, which can be
formulated for QED and hopefully for   QCD,   these approximations should be  related  to  similar
perturbative expressions  for
the  $D$-functions of the NS and SI vector currents.  We discuss  how this CS limit of
$U(1)$ gauge  model
can be specified within perturbation theory.
It is stressed,  that the origin of  this 
limit  differs from a similar limit,
considered in the works  which are devoted
to the searches  for   fixed points
or ``conformally  invariant windows''  in the expressions for the
RG $\beta$-functions of the gauge model under consideration (see e.g.
refs.~\cite{Gardi:1998rf,Shrock:2013uaa}). Its realization also
differs from the proposal of ref.\cite{Armillis:2013wya} to restore the
CS of  renormalized QED by modification of its Lagrangian in an
arbitrary number of dimensions. In our case the mechanism which realizes the CS
is simulating in part the one  that responsible for the CS
in
${\cal N}=4$ SYM theory with  an identical zero  RG $\beta$-function. 
Theoretical applications of the results obtained, including the explanation of the 
cancellations
between SI-type perturbative $\alpha_s^3$ contributions to the GLS sum rule and to the Adler
$D$-function of fermion vector currents ~\cite{Broadhurst:1993ru}  and of the specific  SI $\alpha_s^4$ corrections
to both quantities ~\cite{Baikov:2012zn} are presented. We consider also the
structure of the available $U(1)$ perturbative expansions for the DIS sum rule and for  the Adler functions.
The application of the  certain $\alpha_s^2$ $SU(N_c)$ results for these quantities, which follow from this definition of the
conformally-invariant limit,  are  discussed.

\section{The AVV three-point functions and  the conformal symmetry in
the $U(1)$ and $SU(N_c)$ models.}
\subsection{Basic considerations.}

Let us  study  the three-point functions of eq.~\eqref{tw} and
eq.~\eqref{3} in the conformally invariant  limits of the $U(1)$ and  $SU(N_c)$
gauge models with fermions.
They are realized when the coupling constants and gauge models and the external
SI vector currents $V_{\mu}$ in eq.~\eqref{3} are fixed as the bare unrenormalized quantities. In other
words, in the expressions of the AVV three-point functions of eq.~\eqref{tw} and eq.~\eqref{3} the coupling constants and the concrete
currents are fixed as
$\alpha=\alpha^{B}$, $\alpha_s=\alpha_s^{B}$ and  $A_{\mu}=A_{\mu}^{B}(x)$,
$V_{\mu}(x)=V_{\mu}^{B}(x)$.

In this case two NS vector currents
in the three-point Green
function of eq.~\eqref{tw} and the SI vector current  in the three-point Green
function of eq.~\eqref{3} are conserved by definition,  while the SI
axial-vector operator $A_{\mu}^{B}(x)$ in the three-point Green function of
eq.~\eqref{two} and the SI vector operator $V_{\mu}^{B}(x)$ are not
renormalized by construction
(we recall  that the bare operators do not depend on  any scale or
renormalization constant). As will be discussed below, in 
the Abelian $U(1)$ model with fermions, these requirements can be
formulated in   diagrammatic language and are described by the blocks
of Feynman graphs, where the QED coupling constant $a=\alpha/\pi$  is fixed
and is not renormalized. This  leads to the property $Z_3=1$,
where $Z_3$ is the renormalization constant of the
photon propagator, which is related to
the renormalization
of the QED  coupling constant by     $a=Z_3a^{B}$, where $a=\alpha/\pi$. In this  QED-type  model,
characterized   by the approximation $Z_3=1$,  the
RG $\beta$-function is
identically equal to zero, namely
\begin{equation}
\label{betaZERO}
\beta(a)=
\mu^2\frac{\partial a}{\partial\mu^2}|_{(a)^B~fixed}=\mu^2\frac
{\partial ln Z_3 }{\partial\mu^2}=0~~~.
\end{equation}
This property is equivalent to the existence
of the  CS in this approximation, which
as in the case of ${\cal N}=4$ SYM
theory leads to vanishing of the RG $\beta$-function in all
orders of perturbation theory.

In the case of an   $SU(N_c)$ gauge group, when the renormalized coupling
constant $a_s=\alpha_s/\pi$ can be defined
through the  renormalization of  several vertexes, namely through
 triple-gluon vertex, four-gluon vertex or quark-antiquark-gluon vertex,  it is
unclear how  to formulate
the CS   limit in a manner similar to that discussed above
in the case of the $U(1)$-model, i.e. considering sets of specific Feynman graphs.

Since   the basic requirement $a_s=a_s^{B}$ of this limit
does not depend from the scale parameter the RG $\beta$-function
of this $SU(N_c)$-based model is identically equal to zero in all orders
of perturbation theory
\begin{equation}
\label{betaSUN}
 \beta(a_s)=\mu^2\frac{\partial a_s}{\partial\mu^2}|_{a_s^B~fixed}=0~~.
\end{equation}
The notion of the    CS limit of the   gauge models
turn out to be very  useful for deriving the
relations between concrete analytical scale-independent
perturbative approximations  for  the coefficient functions of   the
Bjp , EJ  and the  GLS    sum rules and
the similar approximations of the   $D$-functions,
constructed from the NS vector currents and the  SI vector currents.
Note. that this consideration
presumes, that the renormalization constant of the
SI axial-vector  current $A_{\mu}(x)=Z^{SI}(a_s)A_{\mu}^{B}(x)$
is fixed as $Z^{SI}=1$.
The basic property, which will be used in the derivation of the
relations mentioned above, is that
in  
the  CS  limit the three-point Green functions of
eq.~\eqref{tw} and eq.~\eqref{3}  have    1-loop expressions,
which are  identical to that of   the AVV Green function in   eq.~\eqref{Tri},
namely
\begin{equation}
\label{tri}
\tilde{T}_{\mu\alpha\beta}^{ab}(p,q)=
\delta^{ab}{\Delta}_{\mu\alpha\beta}^{1-l}(p,q)
\end{equation}
\begin{equation}
\tilde{\tilde{T}}_{\mu\alpha\beta}^{ab}(p,q)=N_F\delta^{ab}{\Delta}_{\mu\alpha\beta}^{1-l}(p,q)
\label{asv}
\end{equation}
where  the  1-loop contributions    ${\Delta}_{\mu\alpha\beta}^{1-l}(p,q)$
on   the r.h.s. of  eq.~\eqref{Tri}, eq.~\eqref{tri} and eq.~\eqref{asv}
are  the same and $N_F$ in eq.~\eqref{asv} appears in view of the fact that the
$SI$ vector current in eq.~\eqref{3} is defined as $V_{\beta}(x)=\overline{\psi}_i(x)\gamma_{\beta}\psi_i(x)$ where
$i$ counts the number of the fermion flavours with identical charges and thus can be re-written as $V_{\beta}(x)=N_F\overline{\psi}(x)\gamma_{\beta}\psi(x)$.
As will be demonstrated, in the CS limit the application
of the operator product expansion (OPE) approach to eq.~\eqref{Tri}, eq.~\eqref{tri} and eq.~\eqref{asv} allow
us to derive relations between the approximations for the coefficient
functions of the DIS sum rules, which will be defined below.

\subsection{The definitions of the perturbative  coefficient functions.}

Within perturbation theory the Bjorken sum rule of   polarized
lepton-nucleon DIS   is defined by
\begin{equation}
\label{Bjp}
S_{Bjp}(a_s(Q^2))=\int_0^1\big(g_1^{lp}(x,Q^2)-g_1^{ln}(x,Q^2)\big)dx=
\frac{1}{6}g_A C_{Bjp}(a_s(Q^2))
\end{equation}
where $g_A$ is the axial nucleon coupling constant.

In the $\overline{MS}$-scheme the  expression for the
polarized Ellis-Jaffe sum rule is
\begin{eqnarray}
 \label{EJp}
EJ^{lp(n)}(Q^2)= \int_0^1 g_1^{lp(n)}(x,Q^2)dx& =&
C_{EJ}^{NS}(a_s(Q^2))(\pm\frac{1}{12}a_3+\frac{1}{36}a_8) \\ \nonumber
&+&
C_{EJ}^{SI}(a_s(Q^2)){\it exp}(\int_{a_s(\mu^2)}^{a_s(Q^2)}
\frac{\gamma_{SI}(x)}{\beta(x)}dx)
\frac{1}{9}\Delta\Sigma(\mu^2)~~~.
\end{eqnarray}
Here  $a_3=\Delta u-\Delta d$=$g_A$,  $a_8=\Delta u+\Delta d- 2 \Delta s$,
$\Delta\Sigma=\Delta u+\Delta d+\Delta s$ and
$\Delta u$, $\Delta d$  and $\Delta s$ are the polarized parton distributions,
while  the subscripts $lp(n)$ labels   the structure functions
$g_1(x,Q^2)$ of   polarized  DIS of  charged leptons ($l$)
on protons ($p$) and neutrons ($n$). Note that
the polarized gluon distribution
$\Delta G$, introduced in QCD  in Refs.~\cite{Efremov:1988zh},~
\cite{Altarelli:1988nr}
does  not contribute to eq.~\eqref{EJp} in   the $\overline{MS}$-scheme
 ~\cite{Zijlstra:1993sh}. The
perturbative expression for the Gross-Llewellyn Smith sum rule of
 neutrino-nucleon DIS can be defined as
\begin{equation}
\label{GLS}
S_{GLS}(a_s)=\frac{1}{2}\int_0^1 F_3^{\nu p+\overline{\nu}p}(x,Q^2)dx=
3C_{GLS}(a_s(Q^2))~~~.
\end{equation}
Considering  the correlator of  two NS
axial-vector currents
\begin{equation}
\label{ANS}
i\int<0|T( A_{\mu}^{a}(x) A_{\rho}^{b}(0))|0>e^{iqx}d^4 x =
\delta^{ab}(q_{\mu}q_{\rho}-g_{{\mu}{\rho}}q^2)
\Pi^{NS}(a_s(\mu^2), Q^2/\mu^2)~~~,
\end{equation}
one can define the Adler
$D$-function of the NS axial-vector currents as
\begin{equation}
\label{DNS}
D_{NS}(a_s(Q^2))=-12\pi^2 Q^2\frac{d \Pi^{NS}(a_s(\mu^2),Q^2/\mu^2)}{d Q^2}=
d_R C_{D}^{NS}(a_s(Q^2))~~~,
\end{equation}
where $d_R$ is the dimension of the quark representation and
 $\mu^2=Q^2$. The   correlator of the two  SI  vector quark currents
has the following transverse form
\begin{equation}
\label{VSI}
i\int<0|T( V_{\mu}(x) V_{\rho}(0))|0>e^{iqx}d^4 x =
(q_{\mu}q_{\rho}-g_{{\mu}{\rho}}q^2)
\Pi^{V}(a_s(\mu^2), Q^2/\mu^2)~~~.
\end{equation}
The corresponding   Adler $D$ can be defined as
\begin{equation}
\label{CdV}
D_{V}(a_s(Q^2))=-12\pi^2 Q^2\frac{d \Pi^{V}(a_s(\mu^2), Q^2/\mu^2)}{d Q^2}=
N_FC_{D}^{V}(a_s(Q^2))~~~,
\end{equation}
where $N_F$ is number of fermion species. Note, that in eq.~\eqref{VSI}  the
expression for the $SI$ fermion vector current is
fixed as $V_{\mu}=\overline{\psi}_i\gamma_{\mu}\psi_i$, which  leads to  the appearance of the factor
$N_F$ in the r.h.s. of eq.~\eqref{CdV}. This particular definition is useful for
considering the cases of both $SU(N_c)$ and  $U(1)$ expressions of eq.~\eqref{CdV} in the same
manner without introducing quark charges $Q_i$ and thus changing the SI
quark vector current in eq.~\eqref{VSI} to the electromagnetic quark current
$J_{\mu}^{EM}=Q_i\overline{\psi}_i\gamma_{\mu}\psi_i$.

The coefficient function of the RG-invariant quantities defined above obey the
following RG equations
\begin{eqnarray}
\label{RG1}
&&\bigg(\mu^2\frac{\partial}{\partial\mu^2}+\beta(a_s)\frac{\partial}
{\partial a_s}\bigg)C_{Bjp}(a_s(\mu^2), Q^2/\mu^2)=0 \\
\label{RG2}
&&\bigg(\mu^2\frac{\partial}{\partial\mu^2}+\beta(a_s)\frac{\partial}
{\partial a_s}\bigg)C_{EJ}^{NS}(a_s(\mu^2), Q^2/\mu^2)=0 \\
\label{RG3}
&&\bigg(\mu^2\frac{\partial}{\partial\mu^2}+\beta(a_s)\frac{\partial}
{\partial a_s}\bigg)C_{GLS}(a_s(\mu^2), Q^2/\mu^2)=0~~\\ \label{RGDNS}
&&\bigg(\mu^2\frac{\partial}{\partial\mu^2}+\beta(a_s)\frac{\partial}
{\partial a_s}\bigg)C_{D}^{NS}(a_s(\mu^2), Q^2/\mu^2)=0~~\\
\label{RGDV}
&&\bigg(\mu^2\frac{\partial}{\partial\mu^2}+\beta(a_s)\frac{\partial}
{\partial a_s}\bigg)C_{D}^{V}(a_s(\mu^2), Q^2/\mu^2)=0~~
\end{eqnarray}
where   $Q^2=-q^2$ is the Euclidean momentum transfer, $\mu^2$ is the
scale parameter of the $\overline{MS}$-scheme. The
coefficient function of the $D$-function of vector currents and of the GLS
sum rule contain the NS and SI-type contributions, namely
\begin{eqnarray}
\label{CDV}
C_{D}^{V}(a_s(\mu^2,Q^2/\mu^2))&=&C_{D}^{NS}(a_s(\mu^2,Q^2/\mu^2))+
N_FC_D^{SI}(a_s(\mu^2,Q^2/\mu^2)); \\
\label{CGLS}
C_{GLS}(a_s(\mu^2,Q^2/\mu^2))&=&C_{GLS}^{NS}(a_s(\mu^2,Q^2/\mu^2))+
N_FC_{GLS}^{SI}(a_s(\mu^2,Q^2/\mu^2))~~.
\end{eqnarray}
where in both  eq.~\eqref{CDV} and eq.~\eqref{CGLS} the SI-type perturbative
corrections   appear   at the $a_s^3$-level (see the
works of refs.\cite{Gorishnii:1990vf}, 
\cite{Surguladze:1990tg},\cite{Gorishnii:1991se},\cite{Chetyrkin:1996ez}
and  ref. \cite{Larin:1991tj} correspondingly).

As was found recently \cite{Larin:2013yba},
the coefficient function of the
Bjp  sum rule also contains  SI-type contributions
\begin{equation}
\label{CBjp}
C_{Bjp}(a_s(\mu^2,Q^2/\mu^2))=C_{Bjp}^{NS}(a_s(\mu^2,Q^2/\mu^2)
+C_{Bjp}^{SI}(a_s(\mu^2,Q^2/\mu^2))
\end{equation}
which appear first at the level of the $a_s^4$ corrections.
It will  be
demonstrated that the NS contributions  to the  coefficient functions of  the DIS  sum rules
coincide in all-orders of perturbation theory, namely that
\begin{equation}
\label{NSsame}
C_{GLS}^{NS}(a_s(\mu^2),Q^2/\mu^2)=C_{Bjp}^{NS}(a_s(\mu^2),Q^2/\mu^2)=C_{EJ}^{NS}(a_s(\mu^2), Q^2/\mu^2)~~.
\end{equation}
These  coefficient functions are  power series in
 $a_s=\alpha_s/\pi$ ,  where    $\alpha_s$ is the  coupling constant of the
$SU(N_c)$ colour gauge theory.
In general the variation of $a_s$ is governed
by  RG   $\beta$-function of the $SU(N_c)$ gauge group,  which  is defined as
\begin{equation}
\label{beta}
\beta(a_s)=
\mu^2\frac{\partial a_s}{\partial\mu^2}|_{(a_s)^B~fixed}=-
\sum_{k\geq 0} \beta_ka_s^{k+2}~~~.
\end{equation}

The RG equation for the SI coefficient function for  the EJ
sum rule contains the anomalous dimension function, namely
\begin{equation}
\bigg(\mu^2\frac{\partial}{\partial\mu^2}+ \beta(a_s)\frac{\partial}
{\partial a_s}+\gamma^{SI}(a_s)\bigg)C_{EJ}^{SI}(a_s(\mu^2),Q^2/\mu^2)=0~~~.
\end{equation}
The anomalous dimension
 of the SI axial current $A_{\mu}$ is defined as
\begin{equation}
\gamma^{SI}(a_s)=\mu^2\frac{\partial ln Z^{SI}(a_s)}{\partial \mu^2}=
 \sum_{l\geq 0} \gamma_l a_s^{l+1}
\end{equation}
where $A_{\mu}=Z^{SI}(a_s)A_{\mu}^{B}$.
This anomalous dimension enters
the four-loop calculations of ref.~\cite{Larin:1997qq},
though its analytical expression in the $\overline{MS}$-scheme  is known
at the three-loop level only   ~\cite{Larin:1994dr} and can be
re-written as
\begin{equation}
\label{gamma}
\gamma^{SI}(a_s)=- \frac{3}{4}C_F T_F N_Fa_s^2
+\bigg(-\frac{71}{48}C_AC_FT_FN_F
+\frac{1}{12}C_F(T_FN_F)^2
+\frac{9}{16}C_F^2(T_FN_F)\bigg)a_s^3 +O(a_s^4)~~~.
\end{equation}
In eq.~\eqref{gamma}   $\gamma_0$ is {\it zero}  due to the fulfilment of the
Ward identities for the SI  axial vector current, $C_F$ and $C_A$ are the
Casimir operators, $N_F$ counts  the number of flavours,  $T_F=1/2$ is the
normalization factor, or the  Dynkin index,  which will be defined below.

\subsection{The definitions of the $SU(N_c)$ and $U(1)$  group weights.}

In order to recall how one can transform perturbative series for the case of the non-Abelian
$SU(N_c)$ gauge model to the case of the Abelian $U(1)$ theory, we follow in this
section the studies, performed in refs.~
\cite{Cvitanovic:1976am},~\cite{vanRitbergen:1997va},~
\cite{vanRitbergen:1998pn},~\cite{Jack:2007ni},
and present the general definitions of the Casimir operators  and the structure constants for the
$SU(N_c)$ and $U(1)$ gauge groups.
The  generators $T^{a}$ of the Lie algebra of the $SU(N_c)$ group  satisfy the
following commutation relations
\begin{equation}
\label{commut}
[T^{a},T^{b}]=if^{abc}T^{c}
\end{equation}
where $f^{abc}$ is the  antisymmetric structure constant.
In a fermion representation, which contains $N_F$  fermions, one has
\begin{equation}
\label{CF}
T^{a}T^{a}=C_F I.
\end{equation}
Here $I$ is the unit matrix and $C_F$ is the quadratic  Casimir operator of the Lie algebra.
The  Casimir operator
$C_A$ of its  adjoint representation
is defined as
\begin{equation}
\label{CA}
f^{acd}f^{bcd}=C_A\delta^{ab}~~~.
\end{equation}
The Dynkin index $T_F$ of the Lie algebra of the  $SU(N_c)$ group
has the following form
\begin{equation}
\label{Dynkin}
Tr[T^{a}T^{b}]=T_F\delta^{ab}~~~.
\end{equation}
Using  eq.~ \eqref{CF} and eq.~ \eqref{CA} one gets
\begin{equation}
\label{dim}
C_Fd_F=T_FN_A
\end{equation}
where $N_A$ is the number of the generators $T^{a}$, and $d_F$ is the dimension of the fundamental  representation $R$ of the $SU(N_c)$
gauge group Lie algebra. The totally symmetric tensor $d_F^{abc}$,   which already appeared
in the discussions presented above, is defined  as
\begin{equation}
\label{dabc}
d_F^{abc}=\frac{1}{2}Tr[T^{a}T^{b}T^{c}+T^{a}T^{c}T^{b}]
\end{equation}
The generators $T^{a}$ of the $SU(N_c)$ colour gauge group are related to   the colour matrixes $\lambda^{a}$ as
$T^{a}=\lambda^{a}/2$. In the fundamental representation $R$ the
concrete analytical expressions of the introduced above
$SU(N_c)$ group characteristics are
\begin{equation}
\label{grf}
C_F=\frac{N_c^2-1}{2N_c}~~,~~C_A=N_c~~~;~~~N_A=N_c^2-1~~~;~~~T_F=\frac{1}{2}~~~,~~~d_F=N_c~~~.
\end{equation}

In this representation the  product of two totally symmetric   tensors  $d_F^{abc}$ equals to
\begin{equation}
d_F^{abc}d_F^{abc}=(\frac{N_ñ^2-4}{N_ñ})(N_ñ^2-1)~~~.
\label{2d}
\end{equation}
It  was  was first  obtained  by the authors of ref. ~\cite{Gorishnii:1990vf}
using the original  method of ref.~\cite{Cvitanovic:1976am}. This expression
was confirmed later on in the detailed work of ref.~\cite{vanRitbergen:1998pn}.

In the case of the Abelian  $U(1)$ gauge group with fermions one has
\begin{equation}
f^{abc}=0~~~~, ~~~~T^{a}= 1, ~~~N_A=1 ~~~.
\end{equation}
Thus, using  \eqref{CF} and \eqref{CA} we get  that in the $U(1)$
model $C_F=1$ and $C_A=0$. Rewriting  \eqref{dim} as
\begin{equation}
T_F=(C_F d_F)/N_A
\label{rgQED}
\end{equation}
taking into account that in QED $d_F=1$,  $N_A=1$   one obtains that for the Lie algebra of the $U(1)$ group
the  Dynkin index is fixed as
\begin{equation}
T_F=1~~~.
\end{equation}
The absence of colour structure in the QED vector
current $J_{\mu}(x)=\overline{\psi}(x)\gamma_{\mu}\psi(x)$ and the application
of eq.~\eqref{dabc} implies that in  QED $d_F^{abc}=1$.

To summarize
\begin{equation}
\label{grfQED}
C_F=1~~,~~C_A=0~~~;~~T_F=1~~~,~~~f^{abc}=0~~~,~~~d_F^{abc}=1~~~,~~~ d_F=1~~~
\end{equation}
are the QED ( or of the Abelian $U(1)$ group) analogs of the group
structures, which appear in the multiloop calculations in the  non-Abelian $SU(N_c)$ group.

\subsection{Specification of the conformally  invariant limit in
perturbative QED}
\label{GQED}

To clarify the  notion  of the CS limit of the  gauge models with fermions to be
used in this work, which does not correspond to the fixed points
or
"conformally  invariant windows"  in   the expressions for the
RG $\beta$-functions of these models, we consider first the case of QED,
and the perturbative quenched QED (pqQED) model in particular.
As  was   recalled above, it was used some time ago in
studies of the finite  QED program.
This model   is described by the set of QED  graphs  without
internal vacuum polarization insertions in  various  multiloop  diagrams.
Among these diagrams is the photon-electron-positron vertex, which  defines
the   renormalization of charge in QED. In the pqQED model
the external photon line  of this photon-electron-positron vertex,
depicted in  Figure 1,    is renormalized  by the multiloop    photon vacuum  
polarization function with the single external  fermion loop only.

\begin{figure}[htbp]
\centering{
\includegraphics[width=.45\textwidth]{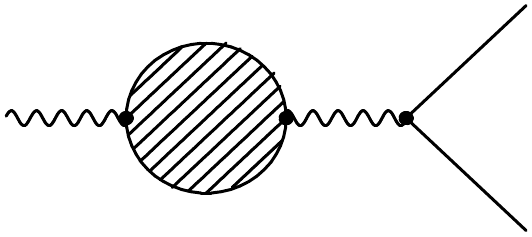}}
\caption{The set of photon vacuum polarization graphs 
which is renormalizing the charge in  the pqQED model as specified below}
\end{figure}

\begin{figure}[htbp]
\centering{
\includegraphics[width=.95\textwidth]{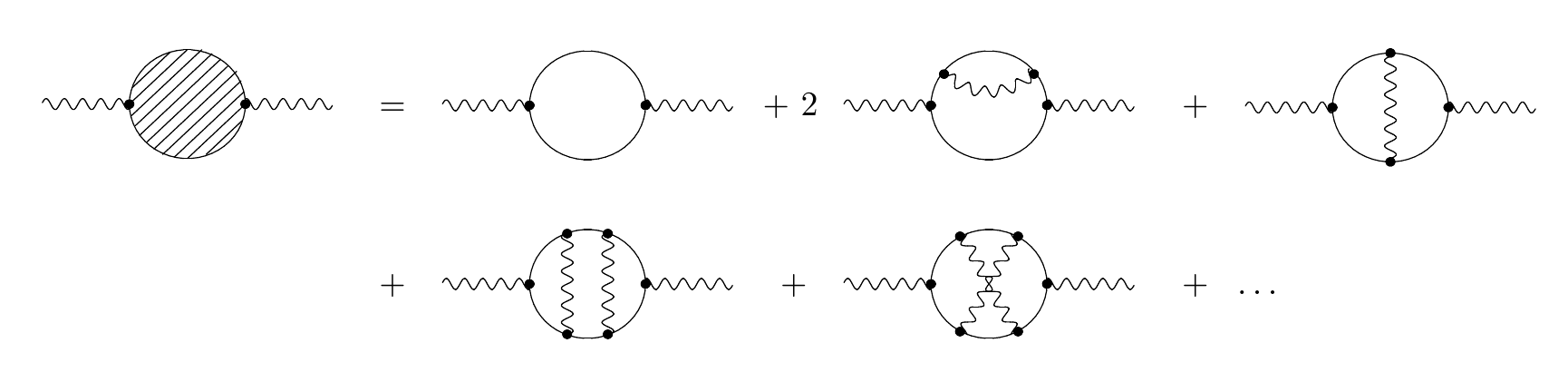}}
\caption{The pqQED approximation of the photon vacuum polarization function which do not contain 
the  graphs 
with lepton-loop insertions into internal photon lines.}
\end{figure}

Of course, there are also the diagram  with the insertion of the            
pqQED approximation of the photon  propagators  into external lepton lines 
of the photon-lepton-lepton vertex and into the vertex itself.  
They  are also                            
contributing into the pqQED approximation of the  renormalization constants     
$Z_2$ and $Z_1$.  But due to the Ward identities $Z_1$ and $Z_2$ are cancelling each other, so only the diagrams          
of Figure 1 are surviving in the definition of the renormalized coupling constant 
of QED and pqQED approximation.

Thus the pqQED approximation of the photon vacuum polarization function is defined by the one-particle irreducible vacuum 
polarization function with the exclusion of diagrams with  lepton-loop inserions into internal 
photon line (see Figure 2). 

In the approximation considered the renormalization constant $Z_3$ of the photon 
propagator
and the related RG $\beta$-function are defined by the photon vacuum 
polarization graphs with one external fermion loop.

Within pqQED model with $N_F=N$ number of leptons
the expression for the $\beta$-function
can be written down as
\begin{equation}
\beta_{pqQED}(a)=\mu^2\frac{\partial a}{\partial \mu^2}=
\sum_{k\geq 0} \beta_k^{[1]}N a^{k+2}
\label{pqQED}
\end{equation}
where $a=\alpha/\pi$, $\alpha$ is the renormalized coupling constant
of the pqQED model,  and the coefficients $\beta_k^{[1]}$ do not depend
from the number of leptons $N$. At the 4-loop level these results
follow the analytical calculations of ref.~\cite{Gorishnii:1990kd},
which were directly   confirmed within the   pqQED model by the calculations
of   ref.~\cite{Broadhurst:1999zi}.

The analytical  result for the  5-loop coefficient
$\beta_4^{[1]}$ was announced   in
ref.~\cite{Baikov:2008cp} and  published later on  in ref.~\cite{Baikov:2010je}
after performing the  calculational
cross-check  outlined  in ref.~\cite{Kataev:2008sk}.  
Note, that since there are  no   sub-divergencies in the
total pqQED  expression for the  photon vacuum polarization function,
the  $\beta$-function of pqQED , as  defined in eq.~\eqref{pqQED},
does not depend on  the choice of the subtractions scheme
to  all orders of perturbation theory.
Therefore,  pqQED approximation is
an example of a  model, where  the remaining  scale dependence manifests itself
in the perturbative expressions  for the related massless  Green
functions only.

Let us now move  one  step further   and define the conformal invlimit of   perturbative QED.
It is realized  when  there is
{\it no  scale} in the theory,  which is
introduced by   charge renormalization. Within the language of 
renormalization
constants this happens when $Z_3=1$. 
This approximation is equivalent to the case when          $a=a^{B}$ 
and $\beta(a)=\beta(a^{B})=0$.

The conditions fixed above define the CS limit of QED considered
in this work
in  diagrammatic language. It  differs from the CS limit
which  is restored in  ref.~\cite{Armillis:2013wya} by the modification
of the QED Lagrangian
in an arbitrary number of  dimensions.

To summarize in our case  the conformally  invariant limit of
perturbative QED exists
\begin{enumerate}
\item
\label{1}
in  the  approximation when only the Feynman diagrams contributing to Green functions
without fermion loop insertions into internal photon lines are considered;
\item
\label{2}
this approximation should be combined with the theoretical  requirement that
in the concrete perturbative  expansions one should use  not the  running coupling,  but the bare
parameter $a=\alpha/\pi$;
\item
\label{2}
the latter property holds when in the photon-lepton-anti-lepton
vertex diagrams  the photon vacuum polarization insertions
(even with single fermion loop) are   not considered.
\end{enumerate}
In other words, in the conformally  invariant limit of QED considered   the photon
vacuum polarization function of  pqQED involves the  two-point Green function
of vector vector currents, while in the vertex Green functions for the   dressed  photon-lepton-
anti-lepton  vertex they are neglected.

In the talk  ~\cite{Kataev:2012kc}
the conformally  invariant  limit of   QED specified above
was used to outline the  derivation of an   all-order  identity between
special contributions to the NS and SI coefficient functions for the
Ellis-Jaffe sum rule  of  polarized lepton-nucleon DIS.
The NS and SI  coefficient functions
$C_{EJ}^{NS}(a_s)$ and $C_{EJ}^{SI}(a_s)$ enter the
operator product expansion expansion of the $T$-product of two
NS  vector currents as
\begin{eqnarray}
\nonumber
i\int T( V_{\alpha}^{a}(x) V_{\beta}^b(0)) e^{ipx} d^4x
|_{p^2\rightarrow\infty}&=&
\delta^{ab}(p_{\alpha}p_{\beta}-g_{\alpha\beta}p^2)
\Pi^{NS}(a_s(\mu^2), P^2/\mu^2) \\ \label{VVOPE}
&+&
 d^{abd}\epsilon_{\alpha\beta\rho\sigma}\frac{p^{\sigma}}{P^2}
C_{EJ}^{NS}(a_s(\mu^2), P^2/\mu^2)~A^{d}_{\rho}(0) \\ \nonumber
&+& \delta^{ab}\epsilon_{\alpha\beta\rho\sigma}
\frac{p^{\sigma}}{P^2}
C_{EJ}^{SI}(a_s(\mu^2),P^2/\mu^2) ~A_{\rho}(0)
\label{OPE}
 \\ \nonumber
&+& {\it higher ~twist~terms}
\end{eqnarray}
where   $P^2=-p^2$ is  the Euclidean transfer
momentum and $\mu^2$ is the renormalization scale parameter.
The second term of the r.h.s. of  eq.~\eqref{VVOPE}  also defines
the   NS part in the   coefficient
function $C_{Bjp}^{NS}(a_s)$ of the Bjp   sum rule
\cite{Gorishnii:1986gn},  which was  introduced above in eq.~\eqref{Bjp}.
Therefore, one has  $C_{EJ}^{NS}(a_s)=C_{Bjp}^{NS}(a_s)$.
The analytical   expression for
the  $a_s^2$- and $a_s^3$-corrections to  $C_{Bjp}(a_s)$
were  analytically evaluated in the $\overline{MS}$-scheme
in the works of  ref.~\cite{Gorishnii:1985xm} and ref.~\cite{Larin:1991tj}
respectively, while the corresponding NS-type
$a_s^4$ contributions
were   evaluated in ref.~\cite{Baikov:2010je}.

To get the non-zero  pqQED analog of the  $O(a_s^3)$ approximation of the   anomalous
dimension $\gamma^{SI}(a_s)$, defined in   eq.~\eqref{gamma},  we use  the discussions
of Section  2.4, fixing   $C_F=1$, $C_A=0$, $T_F=1$ (as follows from eq.~\eqref{grfQED}), and take $N_F^k=0$ for $k\geq 2$ into its available  $SU(N_c)$
expression of  eq. \eqref{gamma}.
In order to put  it to zero  as a  whole  and thus move  to the case of
CS approximation  of
perturbative QED,  it is necessary to add to the requirements (1)-(3),
introduced above while  specifying
the conformal invariant limit of perturbative  QED, the additional condition
of non-renormalization   of the   SI axial-vector  current, i.e.
the condition  $A_{\mu}(x)=A_{\mu}^{B}(x)$.

In this case the correlator of two SI bare axial-vector currents contains
a single external lepton loop and
has the transverse form, namely
\begin{equation}
\label{ASI}
i\int<0|T (A_{\mu}^{B}(x)) (A_{\rho}^{B}(0))|0>e^{iqx}d^4 x =
(g_{{\mu}{\rho}}q^2-q_{\mu} q_{\rho})\Pi^{SI}(a^{B}, Q^2/\mu^2)~~~.
\end{equation}
The corresponding   multiloop approximation of the  formfactor
$\Pi^{SI}(a^{B}, Q^2/\mu^2)$ does  not contain the
diagrams with triangle  contributions to the external  bare vertex, and
therefore the anomalous dimension
$\gamma_{SI}(a)$, discussed in the related  QED studies of
ref.~\cite{Crewther:1973zb}, is absent.
Moreover, in the   conformal invariant limit of  perturbative QED
the formfactor $\Pi^{SI}(a^{B},Q^2/\mu^2)$
coincides with the conformal-invariant
approximation for  the formfactor $\Pi^{NS}(a^{B},Q^2/\mu^2)$ of two
NS axial -vector currents,   which appears in
the QED version of  eq.~\eqref{ASI}.

In the case of perturbative QCD or, more generally, in the  $SU(N_c)$ gauge model
with fermions, the initial CS of the massless quark-parton model can be restored
by fixing $a_s=a_s^{B}$ and $A_{\mu}(x)=A_{\mu}^{B}(x)$.
Using the definitions of eq.~\eqref{beta} and eq.~\eqref{gamma}
one gets  identically zero expressions for the RG functions $\beta(a_s)=0$ and
$\gamma^{SI}(a_s)=0$ and the expression of the transverse expression of the two-point Green
function of the bare axial-vector currents $A_{\mu}^{B}(x)$, which has the form, similar to the
one of \eqref{ASI}.

These pure theoretical  conditions
are useful for the  derivations of  all-order perturbative identities
between the concrete  terms in  the NS and SI coefficient functions of the 
EJ sum rule and
the NS coefficient function of the   Bjp  sum rule and  the GLS  sum rule
 of the
$\nu N$ DIS.
In the $\overline{MS}$-scheme the
total coefficient function of the GLS sum rule $C_{GLS}(a_s)$
 is defined through the OPE of the
NS axial-vector and vector fermion  currents (see e.g.~\cite{Gorishnii:1986gn}) as
\begin{equation}
\label{GLSdef}
i\int T  A_{\mu}^{a}(x)V_{\nu}^{b}(0)e^{ipx} dx=
\delta^{ab}\epsilon_{\mu\nu\alpha\beta}\frac{p^{\beta}}{P^2}
C_{GLS}(a_s(\mu^2,P^2/\mu^2))V_{\alpha}(0)+ {\it higher~twist ~ terms~~}.
\end{equation}
where we follow the notation consistent with the notation for the
Green function of eq.~\eqref{3}. In the next Section we will consider 
theoretical results for the 
DIS sum rules, which follow from the defined in this work CS limit of the 
$SU(N_c)$ and $U(1)$ gauge models.

\section{ Conformal symmetry governed  contributions
to the  coefficient functions of the DIS sum rules in QED and QCD .}
\subsection{The all order perturbative  identities between coefficient functions of DIS sum rules in the  conformal
symmetry limit.}

Let us compare  application of the OPE approach  to the   three  AVV  three-point Green functions of
eq.~\eqref{Tri}, eq.~\eqref{tw}
and eq.~\eqref{3}. In was already discussed in Section 1  that in the CS limit all these three AVV three-point
Green functions are not
renormalized and are expressed through the same one-loop three-point function $\Delta_{\mu\alpha\beta}^{1-l}(p,q)$.
The  results, obtained  in ref.~\cite{Gabadadze:1995ei} in  the kinematic regime $(pq)=0$ (see
ref.~\cite{Gabadadze:1993uc} as well)   demonstrate that  this three-point function can be  expressed through three form-factors:
\begin{eqnarray}
\label{Delta}
\Delta_{\mu\alpha\beta}^{1-l}(p,q)&=& \xi_1^{1-l}(p^2,q^2)
\epsilon_{\mu\alpha\beta\tau}p^{\tau} \\ \nonumber
&+& \xi_2^{1-l}(p^2,q^2)(q_{\alpha}\epsilon_{\mu\beta\rho\tau}p^{\rho}q^{\tau}-
q_{\beta}\epsilon_{\mu\alpha\rho\tau}p^{\rho}q^{\tau}) \\ \nonumber
&+& \xi_3^{1-l}(p^2,q^2)(p_{\alpha}\epsilon_{\mu\beta\rho\tau}p^{\rho}q^{\tau}+
p_{\beta}\epsilon_{\mu\alpha\rho\tau}p^{\rho}q^{\tau})~~~.
\end{eqnarray}

Using   this property and applying the OPE expansion to eq.~\eqref{Tri} and eq.~ \eqref{tw}  in the limit of large $P^2$ and
taking into account the definition of eq. \eqref{OPE},  we get the following two  expressions
\begin{eqnarray}
\label{NS}
d_R^{abc}\Delta_{\mu\alpha\beta}^{1-l}(p,q)&=&i
d_R^{abd}\epsilon_{\alpha\beta\rho\delta}
\frac{p^{\delta}}{P^2}C_{EJ}^{NS}(a_s^{B})
\int<0|T( A_{\mu}^{c}(x)A_{\rho}^{d}(0))|0>e^{iqx}d^4x \\
\label{SI}
\delta^{ab}\Delta_{\mu\alpha\beta}^{1-l}(p,q)&=&i\delta^{ab}
\epsilon_{\alpha\beta\rho\delta}
\frac{p^{\delta}}{P^2}C_{EJ}^{SI}(a_s^{B})
\int<0|T( A_{\mu}^{B}(x)A_{\rho}^{B}(0))|0>e^{iqx}d^4x
\end{eqnarray}
where  the   correlators of the NS and SI axial-vector currents do not contain diagrams with  insertions 
responsible for renormalization of the internal propagators of gauge particles,
and the $SU(N_c)$-group coupling constant $a_s=a_s^{B}$.

Combining now  eq.~\eqref{Delta}, eq.~\eqref{NS} and eq.~\eqref{SI} and 
taking into account   that  in the CS limit the l.h.s. of eqs.~\eqref{NS},~\eqref{SI} do not contain radiative corrections
and that   in the  $SU(N_c)$ gauge model  the correlators of the
NS axial-vector currents $A_{\mu}^{a}(x)$ and of the bare SI axial-vector  currents $A_{\mu}^{B}$ are transverse
(see eq.~\eqref{ANS}   and the $SU(N_c)$-group analog of eq.~\eqref{ASI}) we get
\begin{equation}
\label{xSI}
\xi_2^{1-loop}(q^2,p^2)|_{|p^2|\geq |q^2|>>0}= \frac{1}{p^2}
C_{EJ}^{SI}(a_s^{B})\Pi^{SI}(a_s^{B}, Q^2/\mu^2)~~~.
\end{equation}
This expression is  similar to the CS limit of the expression  obtained in ref.~\cite{Gabadadze:1995ei} in
the NS axial-vector  channel, i.e.
\begin{equation}
\label{XNS}
\xi_2^{1-loop}(q^2,p^2)|_{|p^2|\geq |q^2|>>0}=
\frac{1}{p^2}C_{EJ}^{NS}(a_s^{B})\Pi^{NS}(a_s^{B}, Q^2/\mu^2)~~~.
\end{equation}
Note, that the remaining $Q^2/\mu^2$-dependence comes from the single $log(Q^2/\mu^2)$-terms, which are
related to the overall divergence of the  single-fermion loop approximations of the correlators of NS and
SI axial-vector currents. Taking now the  "weighted" derivatives $-Q^2(d/dQ^2)$ in \eqref{xSI} and
\eqref{XNS} we get the following pair of the   Crewther-type identities
\begin{eqnarray}
\label{DSI}
C_{EJ}^{SI}(a_s)\times C_D^{SI}(a_s)&=&{\bf 1} \\
C_{EJ}^{NS}(a_s)\times C_D^{NS}(a_s)&=&{\bf 1}~~,~
\end{eqnarray}
where $a_s=a_s^{B}$ is considered as the fixed  parameter.
Taking  into account the existence of the property
\begin{equation}
\label{ch}
C_D^{SI}(a_s^{B})\equiv C_{D}^{NS}(a_s^{B})
\end{equation}
which is fulfilled in all orders of perturbation theory, and that $C_{EJ}^{NS}(a_s)=C_{Bjp}^{NS}(a_s)$
(for discussions see Section 2.2 ), we get the following CS based  all-order relation for the
coefficient functions of DIS sum rules
\begin{equation}
\label{DISid}
C_{DIS}^{CS}(a_s^{B}) \equiv  C_{EJ}^{SI}(a_s^{B}) \equiv C_{EJ}^{NS}(a_s^{B})
\equiv C_{Bj}^{NS}(a_s^{B}) \equiv C_{GLS}^{NS}(a_s^{B})~~~~.
\end{equation}
Note, that in the CS limit the ratios of  the corresponding
approximations for the EJ and Bjp sum rules, which are 
determined using their definitions from Section 2.2 and the CS 
identities of eq.~\eqref{DISid}, give us the 
the  following relation
\begin{equation}
\frac{EJ^{lp(n)}(Q^2)}{Bjp(Q^2)}=\pm \frac{1}{2} +\frac{a_8}{6a_3}
+\frac{2\Delta\Sigma}{3a_3}
\end{equation}
where $a_8=3a_3-4D$, $a_3, a_8$ and $\Delta\Sigma$ were  defined
above through the polarized parton distributions  and $D$ is the hyperon decay
constant. These  relations  coincide with the ones 
obtained  in the   massless quark-parton model and
can be rewritten as
\begin{equation}
\frac{EJ^{lp}(Q^2)}{Bjp(Q^2)}=1 +\frac{2(\Delta \Sigma-D)}{3~a_3}
 ~~~~;~~~~\frac{EJ^{ln}(Q^2)}{Bjp(Q^2)}= +\frac{2}{3}\frac{(\Delta \Sigma-D)}{a_3}~~~.
\end{equation}
They lead to the standard   quark-parton model definition of the
Bjp sum rule through  the   EJ sum rules, namely
\begin{equation}
\label{c}
Bjp\equiv EJ^{lp}-EJ^{ln} ~~~.
\end{equation}
This gives us  confidence in the self-consistency of the considerations
presented above.

To get the CS limit expression for the coefficient function of the GLS sum rule, 
we apply the OPE to eq.~\eqref{3} at large $Q^2$. As a result, using the definition of 
eq.~\eqref{GLSdef} in the CS limit of the gauge model under considerations we find that 

\begin{equation} 
\label{relationGLS}
N_F\delta^{ab}\Delta_{\mu\alpha\beta}^{1-l}(p,q)=i\delta^{ab}\epsilon_{\mu\alpha\nu\beta}\frac{q^{\beta}}{Q^2}C_{GLS}(a_s(\mu^2),Q^2/\mu^2)
\int<0|T(V_{\beta}(x)V_{\nu}(0))|0>e^{-ipx}d^4x~~~. 
\end{equation}

The analog of eq.~\eqref{xSI} now reads 
\begin{equation}
\label{xNS}
\xi_2^{1-loop}(q^2,p^2)|_{|q^2|\geq |p^2|>>0}=
\frac{1}{q^2}C_{GLS}(a_s^{B})\Pi^{V}(a_s^{B}, P^2/\mu^2)~~~.
\end{equation}
where $N_F$ on the l.h.s. of eq.~\eqref{relationGLS} comes from the definition of the 
three-point function of eq.~\eqref{3}. Taking the weighted derivative $-P^2(d/dP^2)$ on both sides of eq.~\eqref{xNS}
using the definition of the Adler function of $SI$ vector currents from eq.~\eqref{CDV} we will also use  the discovery 
of finite QED program studies, that the CS limit is also valid in the case of including the SI-type contributions to the 
correlator of SI vector fermion currents which contain two light-by-light scattering subgraphs without internal lepton  loop insertions (see Figure 3). 

\begin{figure}[htbp]
\centering{
\includegraphics[width=.45\textwidth]{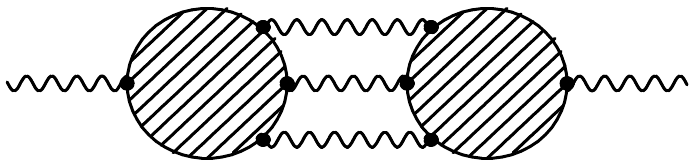}}
\caption{The set of light-by-ligt-type contributions to the
photon vacuum polarization function without internal lepton  loop
insertions.}
\end{figure}
The dashed light-by-light scattering  fermion-loop  subgraphs  of the 
diagrams of  Figure 3 
do not contain diagrams with lepton loops  insertions   into internal 
photons lines (see  Figure 4).   In the approximation we are 
interested  in   these diagrams 
are subtracted from the    one-particle irreducible expressions for the
subset of QED diagrams with four external photon lines, coupled to 
the single lepton loop.    
 
\begin{figure}[htbp]
\centering{
\includegraphics[width=.45\textwidth]{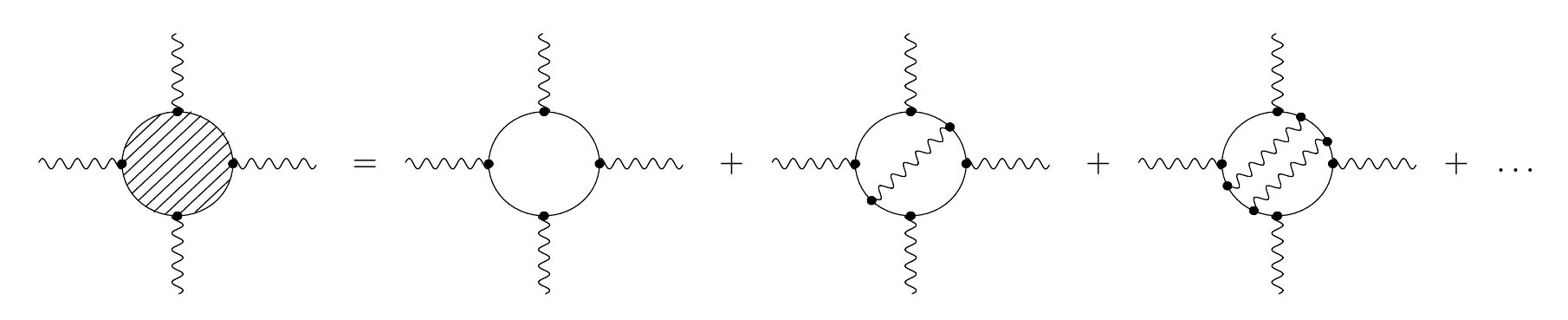}}
\caption{The set of light-by-light scattering graphs without diagrams with  internal
lepton  loop insertions.}
\end{figure}

Taking into account these light-by-light scattering graphs, we arrive at the following analog of the Crewther-type relation between 
the coefficient functions of the GLS sum rule and the Adler function of SI vector currents 
\begin{equation} 
\label{GLSD}
\bigg[\bigg(C_{GLS}^{NS}(a_s^{B})+N_FC_{GLS}^{SI}(a_s^{B})\bigg)\times \bigg(C_D^{NS}(a_s^{B})+N_FC_D^{SI}(a_s^{B})\bigg)\bigg]|_{CS~~limit}=1~~~. 
\end{equation} 
It is valid at the level of taking into account $N_F$-terms for sure, while the cross-check of the $N_F^2$ relation requires the evaluation 
of higher order SI contributions to both $D$-function and to the GLS sum rules, which respects CS limit. At present these require calculations 
going beyond the level of the  analytically  evaluated   $O(a_s^4)$  corrections. Thus, expanding (3.15) in powers of $N_F$ we obtain 
\begin{eqnarray} 
\label{rel1}
C_{GLS}^{NS}(a_s^{B})\times C_{D}^{NS}(a_s^{B})&=&1 \\ \label{rel2}
C_{GLS}^{NS}(a_s^{B})\times C_D^{SI}(a_s^{B})+C_{GLS}^{SI}(a_s^{B})\times C_D^{NS}(a_s^{B})&=&0 
\end{eqnarray}
The expression for eq.~\eqref{rel1} implies that 
\begin{equation} 
C_{GLS}^{NS}(a_s)=1/C_D^{NS}(a_s) 
\end{equation}
and therefore 
\begin{equation}
C_{GLS}^{NS}(a_s^{B})=C_{DIS}^{CS}(a_s^{B}) 
\end{equation} 
where the r.h.s. of eq.~(3.19) is defined by eq.~(3.9). It is identically equal to the 
conformally invariant contributions to the 
SI and NS parts of the $EJ$ sum rule and of the NS contributions to the Bjp sum rule. The expression (3.15) is supporting the 
relation between the $a_s^3N_F$  and the $a_s^4N_FC_F$ corrections to  the SI contribution into the coefficient 
function of the GLS sum rule and into the SI contribution to the coefficient function of the $D$-function of $SI$ vector currents. 
At the $a_s^3$- and $a_s^4$-levels these relations were obtained in ref. \cite{Broadhurst:1993ru} and ref. \cite{Baikov:2010je}
respectively from the results of analytical $a_s^3$ and $a_s^4$ calculations.

\subsection{Concrete analytical and numerical results}

Let us present now some concrete expressions for the scale-independent 
approximations of the several coefficient functions. The first one is the 
expression for the coefficient function of NS $D$-function, obtained in the conformally-invariant 
approximation of QED. It follows from the results of direct analytical 5-loop calculations, 
presented first in the work of   ref. \cite{Baikov:2008cp},  
discussed in detail in the work of ref. \cite{Kataev:2008sk} and published later on  in   ref. \cite{Baikov:2010je}. 
This result has the following form
\begin{eqnarray}
\label{CDQED}
C_{D}^{NS}&=&1+\frac{3}{4}a-\frac{3}{32}a^2-\frac{69}{128}a^3+
\bigg(\frac{4157}{2048}+\frac{3}{8}\zeta_3\bigg)a^4 +O(a^5) \\
& =& 1+0.75a-0.094a^2+0.531a^3+2.481a^4+O(a^5)~~~.
\end{eqnarray} 
In the case of the  $SU(N_c)$ model we will consider in this work   a similar  analytical   $O(a_s^2)$-expression, which 
is known  from the
analysis  of  ref.~\cite{Mikhailov:2004iq}  (see the work of ref.~\cite{Kataev:2010du} as well). It reads  
\begin{eqnarray}
\label{CDQCD}
C_D^{NS}(a_s)&=&1+\frac{3}{4}C_Fa_s +\bigg(-\frac{3}{32}C_F^2+\frac{1}{16}C_FC_A\bigg)a_s^2+0(a_s^3) \\ \label{BLMR}
&=&1+a_s+\frac{1}{12}a_s^2+O(a_s^3) \\ \nonumber
&=&1+a_s+0.083a_s^2+O(a_s^3)~~~~,
\end{eqnarray}
where the numerical expressions for the coefficients are obtained in the case of $SU(3)$ gauge model, namely for $C_F$=4/3 and $C_A$=3. 
Fixing $C_F=1$ and $C_A=0$ in ccordance with the discussions, presented in the Section 3.2, one can reproduce the related   QED results of eq.~\eqref{CDQED} 
from the ones of eq.~\eqref{CDQCD}. In the case of QCD the analytical and numerical expressions for the $O(a_s^2)$-corrections in eq.~\eqref{BLMR} agree 
with the results of application of the BLM scale-fixing approach of ref.~\cite{Brodsky:1982gc}.

In the conformally-invariant limit of perturbative QED   the   results for the coefficient 
functions of   DIS sum rules   can be obtained from the Crewther relations of eq. (3.6) and of eq. (2.7).
In this limit the explicit expression for the NS coefficient function of the Bjorken sum rule was first obtained in 
ref.~\cite{Kataev:2008sk}. Taking it into account we get the following analytical and numerical expressions 
of the considered  in this work identities:  
\begin{eqnarray}
\label{ident}
C_{Bjp}^{NS}(a)&=&C_{EJ}^{NS}(a)=C_{EJ}^{SI}(a)=C_{GLS}^{NS}(a)=1/C_{D}^{NS}(a) \\ \
&=& 1-\frac{3}{4}a+\frac{21}{32}a^2-\frac{3}{128}a^3-
\bigg(\frac{4823}{2048}+\frac{3}{8}\zeta_3\bigg)a^4 +O(a^5) \\ \nonumber 
&=& 1- 0.75a +0.656a^2-0.0234a^3+2.806a^4+O(a^5)~~~.
\end{eqnarray}
The validity of the identity of eq.~(3.25) for the coefficient function $C_{EJ}^{SI}(a)$ was explicitely demonstrated 
in ref. ~\cite{Kataev:2010tm} at the level of $a^3$ corrections. The results of ref.~\cite{Kataev:2010tm} were obtained combining 
the analytical 3-loop expressions, which follow from the results of calculations of refs. \cite{Larin:1991tj}, \cite{Larin:1997qq} 
in the CS limit of QED. 

It will be of interest to check the validity of this  identity in possible direct analytical 4-loop calculations of $C_{EJ}^{SI}(a)$. 

We present here also  the $O(a_s^2)$ CS approximations   for the DIS sum rules in the  the cases  of $SU(N_c)$ and   $SU(3)$ 
models, which result from the considerations of ref.~\cite{Kataev:2010du}:
\begin{eqnarray}
C_{Bjp}^{NS}(a_s)&=&C_{EJ}^{SI}(a_s)=C_{GLS}^{NS}(a_s)=1/C_{D}^{NS}(a_s) \\
&=&1-\frac{3}{4}C_Fa_s+\bigg(\frac{23}{32}C_F^2-\frac{1}{16}
C_FC_A\bigg)a_s^2 
+O(a_s^3)\\ \nonumber
&=&1-a_s+\frac{11}{12}a_s^2+ O(a_s^3) \\ \nonumber
&=&1-a_s+0.917a_s^2+ O(a_s^3)
\end{eqnarray}
The numerical expression for the $O(a_s^2)$-coefficient coincides with
the result obtained in ref.~\cite{Kataev:1992jm} using  the
one-scale $O(a_s^3)$ extension of the BLM approach of ref.~\cite{Brodsky:1982gc}, 
developed in ref.~\cite{Grunberg:1991ac}. 
The  detailed studies 
of the differences  of these results with other extensions of the BLM approach, which 
also have the aim to obtain the expressions for the coefficient functions as a series with scale-independent 
coefficients \cite{Mojaza:2012mf},~\cite{Wu:2013ei},~
\cite{Brodsky:2013vpa} are on the agenda. The method of ref.~\cite{Mikhailov:2004iq}, 
developed for the  formulation of an all-orders extension of the BLM approach, and the new $\overline{MS}$-scheme 
generalization of the Crewther relation \cite{Kataev:2010du}, which presumes the application of the two-fold 
expansion in terms of powers of the $\beta$-function and the coupling constant itself, are quite useful 
for the investigation of this problem.   
   
\section{Conclusions}

In this work using the language of the bare unrenormalized parameters of QED 
and QCD and the properties of the absence of radiative corrections to  
three AVV three-point Green functions in the case when the conformal symmetry 
of $U(1)$ and $SU(N_c)$ gauge models remain valid, we derived three Crewther-type 
relations and the identities between coefficient functions of DIS sum rules. 
We demonstrated how to formulate the conformal-invariant limit of 
perturbative QED on the diagrammatic language and clarified that it is 
possible to fix the similar CS  limit of the $SU(N_c)$ model using the language of bare 
unrenormalized parameters. The concrete checks of the possible analytical evaluations 
of the fourth-order corrections to the SI coefficient functions in the expressions for the EJ 
sum rule within the conformal-invariant expansion in the $U(1)$ model with fermions were outlined. 

The necessity of comparing the results for the coefficient functions of the DIS sum rule 
and the $D$-functions, obtained within the conformal-invariant limit of the $SU(N_c)$
model with fermions, with the the generalizations of the BLM approach were emphasized. 
These generalizations  absorb into the scale of the $\overline{MS}$-scheme 
coupling constant all factors proportional to the coefficients of the $\beta$-function. These 
coefficients are responsible for violation of the CS. 
The CS breaking perturbative effects also manifest themselves in the  $\overline{MS}$-scheme 
generalizations of Crewther relations, discovered , studied , theoretically considered,  proved  and reformulated in the works of 
ref.~\cite{Broadhurst:1993ru},  refs.~\cite{Baikov:2010je,Baikov:2012zn}, ref.~\cite{Gabadadze:1995ei}, refs.\cite{Crewther:1997ux,Braun:2003rp} and 
ref.~\cite{Kataev:2010du}  respectively. The manifestation of these effects through the appearance of loop corrections 
to the AVV three-point functions  starting from the 3-loop level    are supported by the 
manifestation  of a concrete $\beta_0a_s^3$ dependent correction in the explicitly evaluated  3-loop expressions 
for the transverse form-factors of the AVV correlator \cite{Mondejar:2012sz}. More detailed studies of  various expressions for the 
generalized Crewther relations should be quite useful for the task of comparing with existing 
generalizations of the BLM approach. 

Another aim of this   work was to demonstrate that the principles of the CS, which at present 
are widely  considered in the applications of the OPE to the Green functions in ${\cal N}$=4 SUSY 
Yang Mills models, may have an analogy with the 
investigations of the similar problems in the CS limit in both $U(1)$ and $SU(N_c)$ models with fermions.

\acknowledgments

I am grateful   to \fbox{A. P. Bakulev}, D.J.  Broadhurst, S.J.  Brodsky,  
S. V. Mikhailov, D. Mueller,
M.V. Polyakov,  N. Stefanis, K.V. Stepanyntz, O.V. Teryaev and D.V. Shirkov
for their    interest in   subjects related to this work.
I also  wish to thank S.A. Larin for   useful discussions in the process of his  participation at the
preliminary  stage of these studies, which resulted in part in his personal  work of ref.~\cite{Larin:2013yba}.
The work was   supported by   Grants RFBR No. 11-01-00182, 
RFBR No. 11-02-00112 and is done within the scientific program of the 
the Grant NS-2835.2014.2.


\end{document}